\documentclass{iopart}
\usepackage[english]{babel}
\usepackage{indentfirst} 
\usepackage{graphics} 
\usepackage{epstopdf}
\usepackage{epsfig}
\usepackage{float}
\newfloat{sup}{thp}{lop}[section]
\floatname{sup}{Supplementary Material}

\newcommand{\kp}{$k_{P}$ }

\newcommand{\ud}{\,\mathrm{d}}

\begin{document}
\title{ Swarming, Schooling, Milling: Phase diagram of a data-driven fish school model}
\date{}
\author{Daniel S. Calovi$^{1,2}$, Ugo Lopez$^{1,2,3}$, Sandrine Ngo$^4$, Cl{\'e}ment Sire$^{5,6}$, Hugues Chat{\'e}$^4$ and Guy Theraulaz$^{1,2}$}
\address{$^1$ Centre de Recherches sur la Cognition Animale, UMR-CNRS 5169, Universit{\'e} Paul Sabatier, 118 Route de Narbonne, 31062 Toulouse Cedex 4, France}
\address{$^2$ CNRS, Centre de Recherches sur la Cognition Animale, F-31062 Toulouse, France}
\address{$^3$ LAPLACE (Laboratoire Plasma et Conversion d'Energie), Universit{\'e} Paul Sabatier, 118 route de Narbonne, 31062 Toulouse Cedex 9, France}
\address{$^4$ Service de Physique de l'Etat Condens{\'e}, CNRS URA 2464, CEA -- Saclay, 91191 Gif-sur-Yvette, France}
\address{$^5$ Laboratoire de Physique Th{\'e}orique, UMR-CNRS 5152, Universit{\'e} Paul Sabatier, 31062 Toulouse Cedex 4, France }
\address{$^6$ CNRS, Laboratoire de Physique Th{\'e}orique, F-31062 Toulouse, France }
 
\begin{abstract}
We determine the basic phase diagram of the fish school model derived
from data by Gautrais \etal ({\it PLoS Comp. Biol.} {\bf 8}, e1002678
(2012) \cite{Gautrais2012}), exploring its parameter space beyond the
parameter values determined experimentally on groups of barred
flagtails ({\it Kuhlia mugil}) swimming in a shallow tank. A modified
model is studied alongside the original one, in which an additional
frontal preference is introduced in the stimulus/response function to
account for the angular weighting of interactions. Our study, mostly
limited to groups of moderate size (in the order of 100 individuals),
focused not only on the transition to schooling induced by increasing
the swimming speed, but also on the conditions under which a school can
exhibit milling dynamics and the corresponding behavioral transitions.
We show the existence of a transition region between milling and
schooling, in which the school exhibits multistability and intermittency
between schooling and milling for the same combination of
individuals parameters. We also show that milling does not occur for
arbitrarily large groups, mainly due to a distance dependence
interaction of the model and information propagation delays in the
school, which cause conflicting reactions for large groups.
We finally discuss the biological significance of our
findings, especially the dependence of behavioural transitions on
social interactions, which were reported by Gautrais \etal to be
adaptive in the experimental conditions.
\end{abstract}

\maketitle

\section{Introduction}

Transitions between different types of collective behaviour play a
major role in the adaptiveness of animal groups \cite{Couzin2003,
  Sumpter2010, Sole2011}. These transitions are commonly observed when
animal groups shift from one collective behaviour to another either
spontaneously or in response to a threat. For instance in fish
schools, individuals may adopt different spatial patterns when they
are travelling, feeding or displaying defensive behaviours
\cite{Radakov1973, Pitcher1993, Parrish1999, Freon1999, Parrish2002,
  Becco2006}. It is commonly observed that periods of swarming
associated with feeding behaviour, where the group remains cohesive
without being polarized, are often interspersed with brief periods of
schooling during which fish search for food and move from
one place to another \cite{Radakov1973}. The actual mechanisms and
behavioural rules that trigger such transitions are still poorly
understood.

Several models of collective motion have been introduced in this
context.  For instance, in the Aoki-Couzin model \cite{Aoki1982,
  Couzin2002}, sharp transitions in collective behaviour are observed
for small parameter changes. In this zonal model, slight variations in
the width of the alignment zone which controls the alignment behaviour
of fish to their neighbours can yield drastic changes of the structure
and polarisation of the school, such as transitions between schooling,
milling (collective vortex formation) and swarming. In the physics
literature, transitions between different types of collective
behaviour are often described as phase transitions. A well studied
case is the transition to orientational order observed in the
popular Vicsek model \cite{Vicsek1995,Czirok1997, Chate2008}.
Similar transitions have been described in self-propelled particle
models by varying the value of parameters such as the blind angle
\cite{Newman2008}, individual speed \cite{Mishra2012}, or a ``strategy
parameter'' \cite{Szabo2008}, which ponders a behavioural compromise
between aligning with neighbours and reacting to their direction
changes.

All the above models suffer from a lack of experimental validation,
having often been developed in a very general context not particularly
in relation to experiments/observations.  However a methodology to
build models for animal collective motion from the quantitative
analysis of trajectories in groups of increasing sizes has been
recently proposed \cite{Gautrais2012}.  Using video tracking of groups
of barred flagtails ({\it Kuhlia mugil}) in a shallow tank, the
stimulus/response function governing an individual's moving decisions
in response to the position and orientation of neighbours was
extracted from two-dimensional trajectory data. It was found that a
gradual weighting between alignment (dominant at short distances) and
attraction (dominant at large distances) best accounted for the
data. It was also found that the parameters governing these two
interactions depend on the mean speed of the fish, leading to an
increase in group polarisation with swimming speed, a direct
consequence of the predominance of alignment at high speed.

Here we investigate the basic phase diagram of the Gautrais \etal 2012
model, exploring its parameter space beyond the parameter values
determined from the data. This is done in order to explore the full
range of collective behaviours that could be displayed by the
model. Our study is limited to groups of moderate size (in the order
of 100 individuals) and focuses, beyond the transition to schooling
induced by increasing the swimming speed, on the conditions under
which a school can exhibit milling dynamics.  The dedimentionalized
model reveals that changes in swimming speed are equivalent to
changing the two social interaction parameters while maintaining the
noise constant. Exploring the parameter plane defined in this way, we
find four regions with distinctive collective behaviour. We show in
particular that in this model, as in most others, milling does not
occur for arbitrarily large groups.

\section{Model}
\label{SecModel}

Let us first recall the findings of Gautrais \etal \cite{Gautrais2009,
  Gautrais2012}.  Barred flagtails ({\it Kuhlia mugil}) in a circular
4 meters diameter tank with a depth of 1.2 meters, were video-recorded
from above for a few minutes in groups of one to 30 individuals. In
this quasi 2D geometry, very few ``crossing'' events (with one fish
passing under another) were observed.  At least 5 replicates were
performed for each group size. In each experiment, fish synchronised
their (mean) swimming speed, and their instantaneous speed (tangential
velocity), although fluctuating by approximately 10-20\% around its
mean, was found uncorrelated to their turning speed (angular
velocity).  This led to the representation of a the single fish
behaviour as a two-dimensional smooth random walk in which an
Ornstein-Uhlenbeck process \cite{Uhlenbeck1930} acts on the fish
angular velocity $\omega$.  The angular velocity is mathematically
described by the stochastic differential equation:
\begin{equation}
\ud \omega(t) = - v \left[ \frac{\ud t}{\xi} \left(\omega(t) -
\omega^{*}(t)\right) + \sigma \ud W \right], 
\label{AngularSpeed}
\end{equation}
where $v$ is the (constant) swimming speed, $\ud W$ refers to a
standard Wiener process, $\xi=0.024$ m is a characteristic persistence
length and $\sigma=28.9$ m$^{-1}$s$^{-1/2}$ controls the noise
intensity, where both $\xi$ and $\sigma$ have been estimated from
experimental data on barred flagtail trajectories.

\begin{figure}[!htb]
\begin{center}
 \includegraphics[width=0.9\linewidth]{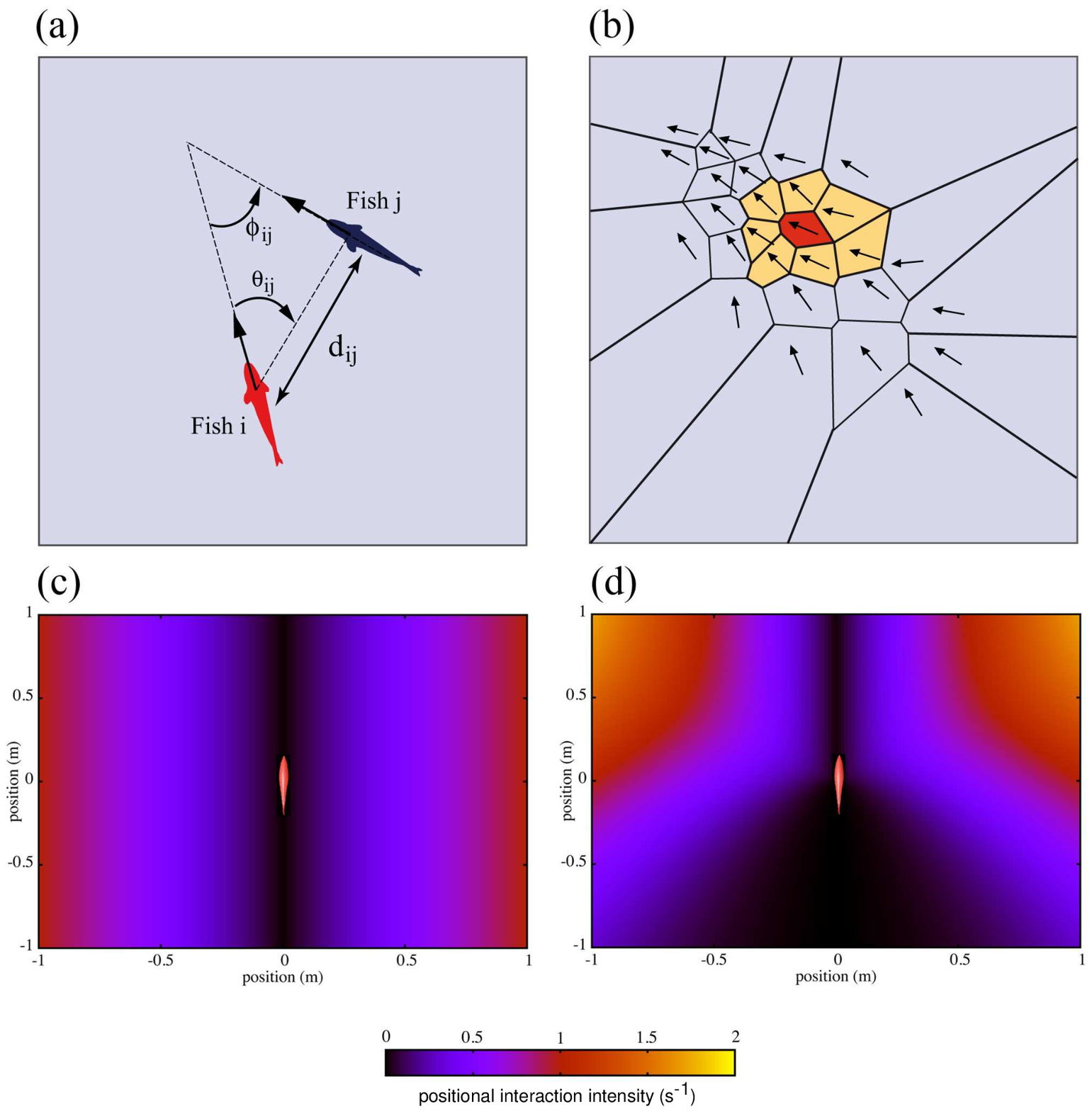}
\caption{Fish interactions graphical representation used in the
  model. $(a)$ The distance $d_{ij}$ of fish $j$ from fish $i$;
  $\phi_{ij}$ is the relative orientation heading of fish $j$ compared
  to fish $i$; $\theta_{ij}$ is the angle between the angular position
  of fish $j$ with respect to fish $i$. $(b)$ illustration of the
  Voronoi neighbourhood, where arrows indicate the fish headings of
  the focal fish (red region) to his first Voronoi neighbours (yellow
  regions). $(c)$ Positional interaction intensity as proposed by
  Gautrais {\it et al} $(d)$ Positional interaction intensity as given
  by the new angular preference.}
\label{Interactions}
\end{center}
\end{figure}

In the above equation, $\omega^{*}(t)$ is the stimulus/response
function based on the individuals response to the proximity of the
tank wall ---irrelevant in the following as we shall consider groups
evolving in an infinite domain--- and on the reaction to the
neighbouring fish orientation and distance.  It was found that for
groups of two fish, these social interactions are well described by a
linear superposition of alignment and attraction with weights
depending on $d_{ij}$, the distance between the two fish.  It was
found further that in groups of more than two fish, many-body
interactions could be safely approximated by the normalized sum of pair
interactions with those individuals forming the first shell of Voronoi
neighbours of the focal fish (\fref{Interactions}). Mathematically,
one has:
\begin{equation}
\omega^{*} =\frac{1}{N_i}\sum_{j \in V_i}\left[k_V \, v \, \sin
  \phi_{ij} + k_P \, d_{ij} \, \sin \theta_{ij}\right]
\label{Omega*}
\end{equation}
where $V_i$ is the Voronoi neighbourhood of fish $i$ containing
$N_i+1$ individuals, $\phi_{ij}=\phi_j -\phi_i$ is the angle between
the orientational headings of fish $i$ and $j$, and $\theta_{ij}$ is
the angle between the heading of fish $i$ and the vector linking fish
$i$ to fish $j$ (\fref{Interactions}).  The functional forms involving
the sine function, chosen for simplicity, were also validated by the
data. They ensure that interactions vanish when the fish are already
aligned or in front of each other as can be seen in
\fref{Interactions}$(c)$.

Note there is a speed dependence with the alignment, and hence a
tendency to produce polarised groups at large speeds.  A similar
positive correlation between group speed and polarity has also been
found in groups of giant Danios ({\it Devario aequipinnatus}) and
golden shiners ({\it Notemigonus crysoleucas})
\cite{Viscido2004,Tunstrom2013}.

Note also that the strength of the attractive positional interaction,
scaled by \kp, increases linearly with $d_{ij}$, which is rather
unrealistic: if two fish were very far apart, they would turn at
extremely large speeds towards each other. Moreover, attraction is of
the same amplitude whether fish $j$ is in front or behind fish $i$,
which may also look unrealistic, as one might expect fish $i$ to be
less ``aware'' and attracted to fish $j$ if $j$ is behind
$i$. Nevertheless this form of attraction interaction was found to
account well for the data, because (i) in the experimental tank fish
never go far away from each other; (ii) data were too scarce to allow
for detecting a frontal preference. In the following, we restrict
ourselves to moderate-size groups, so that the fish inter distances
remain rather small, and we could thus keep the unbounded expression
of the positional interaction strength. However, we will consider the
possibility of an angular weighting of the interactions in the form of
an extra multiplicative factor
\begin{equation}
\Omega_{ij} =1 +\cos( \theta_{ij} )
\end{equation}
to the interaction term $\omega^{*} (t)$, so that \eref{AngularSpeed}
is rewritten as:
\begin{equation}
\ud \omega(t) = - v \left[ \frac{\ud t}{\xi} \left(\omega(t) -
  \Omega_{ij} \omega^{*}(t)\right) + \sigma \ud W \right],
\label{NewAngularSpeed}
\end{equation} 
The function $\Omega_{ij}$ was designed to ensure a frontal preference
and some kind of rear blind angle.  Its specific form (which can be
seen in \fref{Interactions}$(d)$) was chosen for its simplicity and its
unit angular average.

In the following, we perform numerical simulations of the model,
varying its key parameters: the swimming speed $v$, and $k_P$ and
$k_V$, the behavioural parameters controlling attraction and
alignment. Most of our study is restricted to moderate-size groups of
$N=100$ fish. However, we also present results as a function of group
size when investigating the robustness of milling (\sref{SecSize}).
Typically, a set of 100 random initial conditions is simulated over a
period of 1000 seconds, where the first half of each series was
discarded to account for transient behaviour.

The polarisation of the school is quantified by the global polar order
parameter
\begin{equation}
P= \frac{1}{N} \left|\sum_{i=1}^{N} \frac{\vec{v}_i}{v}\right|
\label{P}
\end{equation}
which reaches order 1 values for strongly polarised, schooling groups.
Milling behaviour is detected via the global normalised angular momentum
\begin{equation}
M= \frac{1}{N} \left| \sum_{i=1}^{N}{\frac{ \vec{r}_i \times
    \vec{v}_i }{|\vec{r}_i| v}}\right|
\end{equation}
which will tend towards unity when the school is rotating in a single
vortex.  As we shall see, the two scalar, positive, quantities $P$ and
$M$ often play symmetric roles in the ordered phases of the groups: a
schooling group will show large $P$ and negligible $M$, and the
opposite is true for a milling group.

\section{Speed-induced transition to schooling}
\label{SecUgo}

\begin{figure}[!htb]
\begin{center}
\includegraphics[height=0.8\textheight]{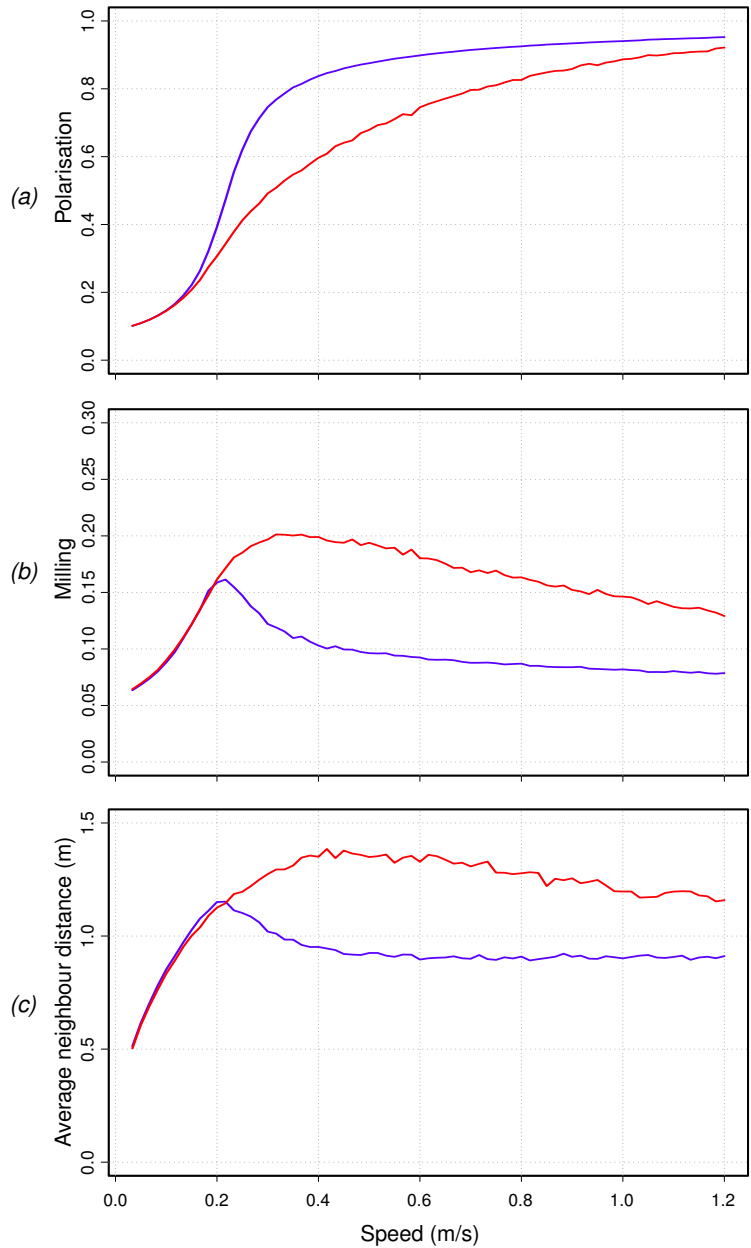}
\caption{$(a)$ Displays the polarisation parameter as a function of
  the speed $v$, for simulations with the original model (blue) and
  the new angular preference (red), where a collective state
  transition to schooling can be seen in both cases, although more
  abruptly in the original model (blue line). $(b)$ milling parameter
  for both simulations (blue original model and red with the new
  angular preference) as a function of the speed. One observes that in
  the transition zone ($v \approx$ 0.2 m/s) there is an increase in
  milling values in the new angular preference simulations
  (red). $(c)$ average distance to the neighbours for both sets of
  simulations. We performed 10 replicates for each different speed, and all simulations were had $10^7$ iterations, where we discarded the first half of data, and used only one datapoint for every 10 iterations.}
\label{Ugo}
\end{center}
\end{figure}

We first report the basic transition from swarming to schooling
observed when increasing the swimming speed of individuals. Already
observed in \cite{Gautrais2012}, it is the natural consequence of the
predominance of the alignment interaction at high speeds.  Here, we
confirm the existence of the transition in the absence of walls
confining the group.  Figure \ref{Ugo} summarises our results for
$N=100$ individuals, using the parameter values determined from the
experimental data obtained in smaller groups ( $k_V $ = 2.7 m$^{-1}$,
$k_P$ = 0.41 m$^{-1}$s$^{-1} $, $\xi$ = 0.024 m and $\sigma$ =28.9
m$^{-1}$s$^{-1/2}$), with and without the angular preference function
$\Omega$.

We find that the transition to schooling occurs at a speed range of
zero to four body lengths per second, a reasonable one considering the
barred flagtail swimming ability (\fref{Ugo}$(a)$). The transition is
smoother with angular preference (red curves) and consequently
requires larger speed values to reach similar results as the original
model. Even though no pronouced milling is found in either cases of 
figure \ref{Ugo}$(b)$, a more pronounced result is achieved with the
angular preference, which seems to directly influence the school
distance to the centroid on figure \ref{Ugo}$(c)$.

\section{Dedimensionalisation of the model}

In order to get more insight into how speed affects the relative
importance of noise, directional and positional behavioural reactions,
we rewrite the model in a dedimensionalized form. After a straightforward
manipulation we are left with only three independent parameters:
\begin{equation}
\alpha=\xi^{3}\sigma^2/v,\;\;\; \beta = ( k_V v) /( \xi^2 \sigma^2),\;\;\; {\rm and} \;\;
\gamma = (k_P v)/(\xi^4 \sigma^4)\;.
\end{equation}
The time step $\ud \tilde{t}=\ud
t/(\xi^2 \sigma^2)$ has no influence on the simulations provided it is
small enough. For all simulations we used a $\ud \tilde{t} =0.1
\alpha $ for $\alpha < 0.1$ or $\ud \tilde{t} =0.01$ for the
remaining cases. The full set of dedimensionalised equations reads:
\begin{equation}
\alpha\ud \tilde{\omega}(\tilde{t}) = \ud \tilde{t} \left(\tilde{\omega}(\tilde{t}) -
\tilde{\omega}^*(\tilde{t})\right) +  \ud W ,
\label{Adim}
\end{equation}
where $\ud W$ refers to a standard Wiener process. The term
$\tilde{\omega}^*$ rewrites as:
\begin{equation}
\tilde{\omega}^* =\frac{1}{N_i}\sum_{j \in V_i}\left[\beta \sin
\phi_{ij} + \gamma \, \tilde{d}_{ij} \, \sin \theta_{ij}\right].
\label{Omegand*}
\end{equation}
Note that in our equivalent system, time is now counted in units of
a typical time $\tilde{x}=v(\xi\sigma)^{-2}$, defined with the
correlation length $\xi$ and noise amplitude $\sigma$. The distance is
expressed in units of a typical length $\tilde{x}=v(\xi\sigma)^{-2}$,
which is simply the displacement of a fish during time $\tilde{t}$
with speed/velocity $v$. The angular speed and thus the evolution of
the fish orientation are determined by the relative weight of
equivalent alignment ($\beta$) and positional ($\gamma$) intensity
over the equivalent angular inertia term $\alpha$.


The expressions of the dedimensionalized coefficients show that
increasing speed, while keeping other parameters constant, leads to an
increase in the equivalent positional and alignment interactions. At
the same time it reduces the angular inertia while keeping the
equivalent noise amplitude/intensity constant. The speed induced
transition can thus be seen as a competition between noise and social
interactions.

Hereafter we perform an extensive study of the dedimensionalised
version of the model, with particular attention paid to the emergence
of significant milling.

\section{Phase diagram in the space of behavioural parameters}
\label{SecMilling}

\begin{figure}[!htb]
\begin{center}
 \includegraphics[height=0.75\textheight]{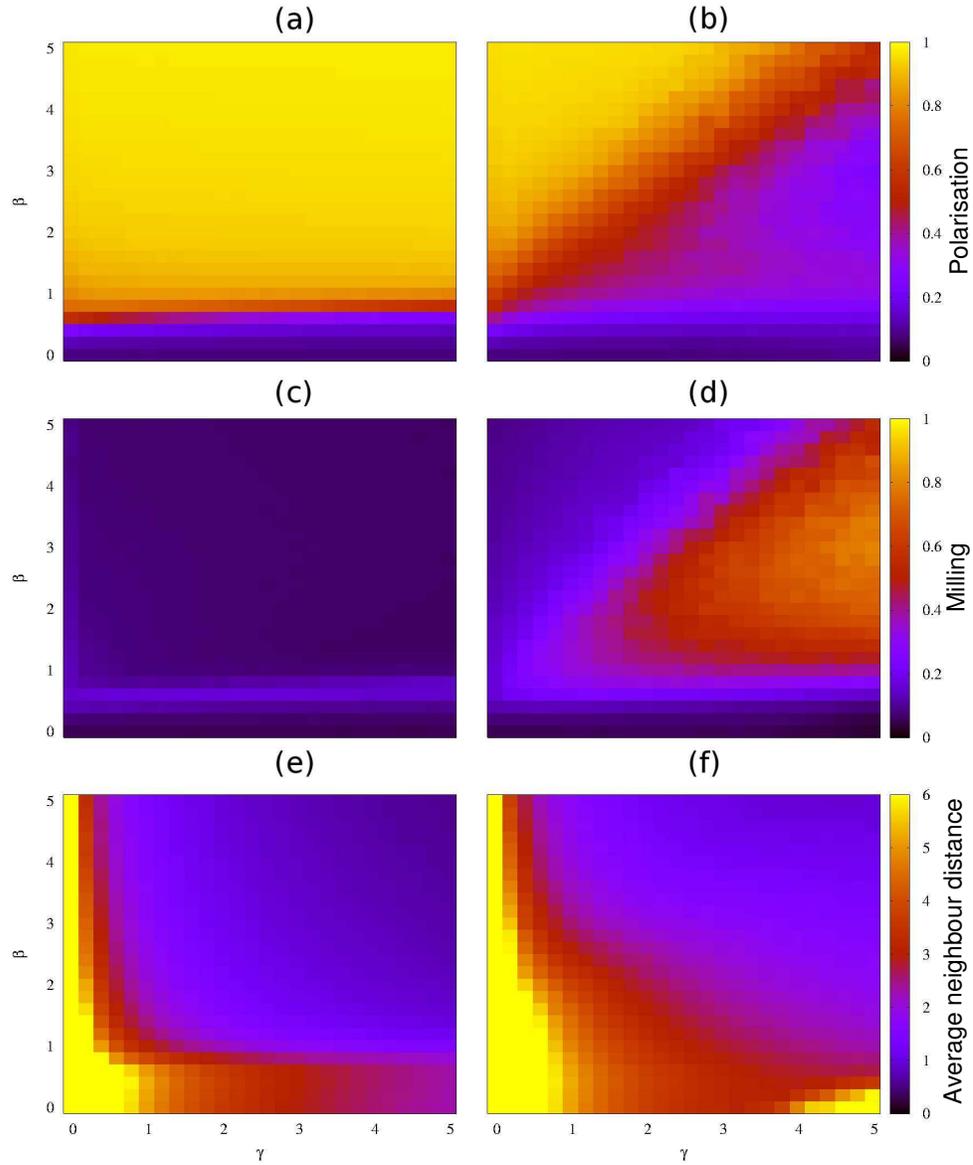}
\caption{Statistics for simulations made for different alignment and
  positional parameters ($\beta$ and $\gamma$ respectively), both in
  the interval [0,5] with an increment of 0.2 (in both $\beta$ and
  $\gamma$ units respectively). Simulations were made with an angular
  inertia term of $\alpha \approx 0.014$ (equivalent to a speed $v=0.8
  {\rm m/s}$). The colour code bar on the right represents the values
  for the respective statistics on the left (polarisation, milling and
  average distance to the neighbours). The left column represents
  different statistics for the original model as seen in
  \eref{AngularSpeed}, while the right column graphs represent
  simulations with the new angular preference as seen in
  \eref{NewAngularSpeed}. The statistics here represented are: $(a)$
  and $(b)$ polarisation parameter; $(c)$ and $(d)$ milling parameter;
  $(e)$ and $(f)$ the average neighbour distance. Each data point is
  the result of an average over 100 different simulations, where the
  first half of each simulation was discarded to avoid transional
  behaviours, resulting in a total of 2$\times 10^6$ data points for
  every parameter configuration.}
\label{v04F0F2}
\end{center}
\end{figure}

 We explored the parameter plane formed by coefficients $\beta$ and
 $\gamma$ in the domain $[0,5]\times[0,5]$ including the
 experimentally-determined values $k_V $ = 2.7 m$^{-1}$ ($\beta\approx
 2.24$) and $k_P$ = 0.41 m$^{-1}$s$^{-1} $ ($\gamma\approx 0.71$) by
 steps of 0.2, yielding the $26\times 26$ grid of results shown in
 \fref{v04F0F2}. The angular inertia term was set at $\alpha\approx
 0.014$, equivalent to the experimentally reasonable value $v=0.8{\rm
   m/s}$.

Two cases are presented, without and with the angular weighting
function $\Omega$ (left and right panels respectively). In both cases,
a transition to schooling is observed as $\beta$ is increased (figures
\ref{v04F0F2}$(a$) and \ref{v04F0F2}$(b)$). Such a transition is
virtually independent of $\gamma$ considering the original model
(figure \ref{v04F0F2}$(a)$) while presenting a functional form for
simulations using the new angular dependence (figure \ref{v04F0F2}).
  
A strong qualitative difference is nevertheless observed between the
left and the right panels in figures \ref{v04F0F2}$(c)$ and
\ref{v04F0F2}$(d)$. With the angular dependence $\Omega$ (right
panels), outright schooling is preceded by a large region where
milling is observed (\ref{v04F0F2}$(d)$), whereas no significant
milling occurs without angular weighting (\ref{v04F0F2}$(c)$).  This
may not seem too surprising given that $\Omega$ favours the formation
of (local) files in which fish follow those in front of them.
	
	
Lastly, we observe a peculiar property in figures \ref{v04F0F2}$(e)$
and \ref{v04F0F2}$(f)$, which show the average distance to neighbours.
With the angular preference (\fref{v04F0F2}$(f)$), the region of low
$\beta$ exhibits a weak ridge bordering the swarming region ($\gamma$
and $\beta \approx 0$) where the average distance is expected to
diverge. This ridge extends to high $\gamma$ values for which one
would infer small-size schools and shorter neighbour distances. One
should be aware that the colour code presented on these last two
figures interpret values above the range [0,6] all as having the same
colour (yellow). This was introduced since a divergence occurs in the
swarming region, but this unfortunately masks secondary ridges, such
as the one found in \fref{v04F0F2}$(f)$. Similar regions may be found
in figure \ref{v04F0F2}$(e)$, although masked by the swarming region
divergence in this visualization.

The explanation, despite being initially counter intuitive, is rather
simple. Despite the $\gamma$-induced tendency to stay together, a low
$\beta$ means that fish have a lower tendency to disturb their
alignment to match that of their neighbours, meaning that fish will
rely mostly on the positional interaction. This implies that if a
neighbour is directly ahead or behind the focal fish, there is no need
to change the direction of motion. Also, both cases (with or without
the new angular preference) have a zero interaction for anti-parallel
neighbours. This means that when $\gamma$ dominates the interactions,
the stable fish configuration is a line. Having noise in
\eref{AngularSpeed} or \eref{NewAngularSpeed} leads to fish organizing
themselves in a quasi-unidimensional configuration, and the fish in
front, not having additional neighbours on their sides, will
eventually turn due to noise. After a slight turn, this fish will now
have new Voronoi neighbours, with fish previously located behind it
now located slightly to its side, and make a 180 degree turn to
recover the optimal configuration. This mechanism leads fish to
organize themselves in two columns going in opposite directions
connected at the end of the school by this ``U turn'' (see snapshot in
\fref{Phases}$(c)$, panel III).

To visualize the above results in a synthetic way, we took advantage
of the fact that the main behavioural regions apparent in figures
\ref{v04F0F2}$(b)$, \ref{v04F0F2}$(d)$ and \ref{v04F0F2}$(f)$ are
distinguished by mutually-exclusive quantities: when schooling is
strong, milling is weak and the distance to neighbours ($D$) is small;
when milling is strong, schooling is weak, and $D$ is small; when $D$
is larger, both schooling and milling are weak.  This mutual exclusion
of schooling and milling is clear in the time series shown in
\fref{Phases}$(a)$.  We thus constructed the following composite
order parameter:
\begin{equation}
S(P,M,D)= P+ M\exp[i 2\pi/3]+ D\exp[i 4\pi/3]
\end{equation}
which takes complex values such that the phase codes for the behaviour
and the modulus for the intensity of this behaviour, where the average
distance was normalised according to the largest value observed. 
As a result, the information of the right panels of \fref{v04F0F2}
is represented synthetically in \fref{Phases}$(b)$.
Three distinct regions are clearly
apparent as each is represented in a different colour (red for
polarisation, blue for milling and green for average neighbour
distance). Typical configurations of the group corresponding to the
time-series of \fref{Phases}$(a)$ are shown in \fref{Phases}$(c)
$. (See also the movies in the supplementary material \ref{RegionI} to
\ref{RegionIII} for simulations of these 4 behaviours.)

Regions I and II refer to the schooling and milling states
respectively.  Region III refers to the winding (line configuration)
state mentioned previously. It could be argued that regions II and III
are similar, having only changed the width/length ratio. Nevertheless
region III has very weak rotational behaviour at odds with full blown
milling.  Note finally the intermittence between schooling and milling
in the transition region between regions I and II.

Finally, we come back to the effect of the swimming speed. The above
analysis in the $(\beta, \gamma)$ plane was repeated at different
values of $\alpha$ corresponding to speed values of 0.4, 0.8 and 1.2
m/s, building a three-dimensional phase diagram. In
\fref{Trans} we show that these transition lines superimpose onto a
master curve delimiting the schooling and milling zones.  Thus, no
qualitative changes were observed for these values of $\alpha$,
indicating the robustness of the phase diagram in \fref{Phases} for
speeds in the experimental range. We studied this phase boundary
marking the transition, via intermittent regimes, between schooling
and milling more quantitatively.  Defining the transition points as
those where both the polarisation and milling parameters spend more
than 40 percent of the time above the threshold value of 0.8, we find
that the transition line, for all values of $v$ studied, can be fitted
by the simple functional form
\begin{equation}
\beta= A \sqrt{\gamma}+ B
\end{equation}
where $A$ and $B$ take values independent of $\alpha$ in the ranges
studied (\fref{Trans}).

Although we do not have a quantitative explanation for the above
functional form, these fits indicate the possibility of a rather
simple theory accounting for the full phase diagram of our
model. Meanwhile a qualitative explanation for the dependence on
$\gamma$ is given by the need to counteract the global polarisation
tendency given by $\beta$, while maintaining a local one. Also, figure
\ref{v04F0F2}$(a)$ indicates minimum $\beta \times \gamma$
values to overcome the noise. In the supplementary material
there are 2 videos (\ref{beta} and \ref{gamma}) of the two dimensional
histogram evolution for the schooling and milling parameters as we
change $\beta$ or $\gamma$ while maintaining the other one constant (
$ \gamma=17.28$ and $\beta=13.30$), meaning we have respectively
vertical and horizontal transition cross-sections in \fref{Trans}.

\section{Group-size-induced transition}
\label{SecSize}


In this section, we investigate the influence of group size on the
robustness of both milling and schooling behaviour in our model using
the angular weighting function. Apart from the general theoretical
context mentioned above, this is also of direct relevance for fish, as
it has been reported recently that in golden shiners increasing group
size significantly increases the amount of time spent in a milling
state \cite{Tunstrom2013}.

\begin{figure}[!htb]
\begin{center}
 \includegraphics[width=0.9\linewidth]{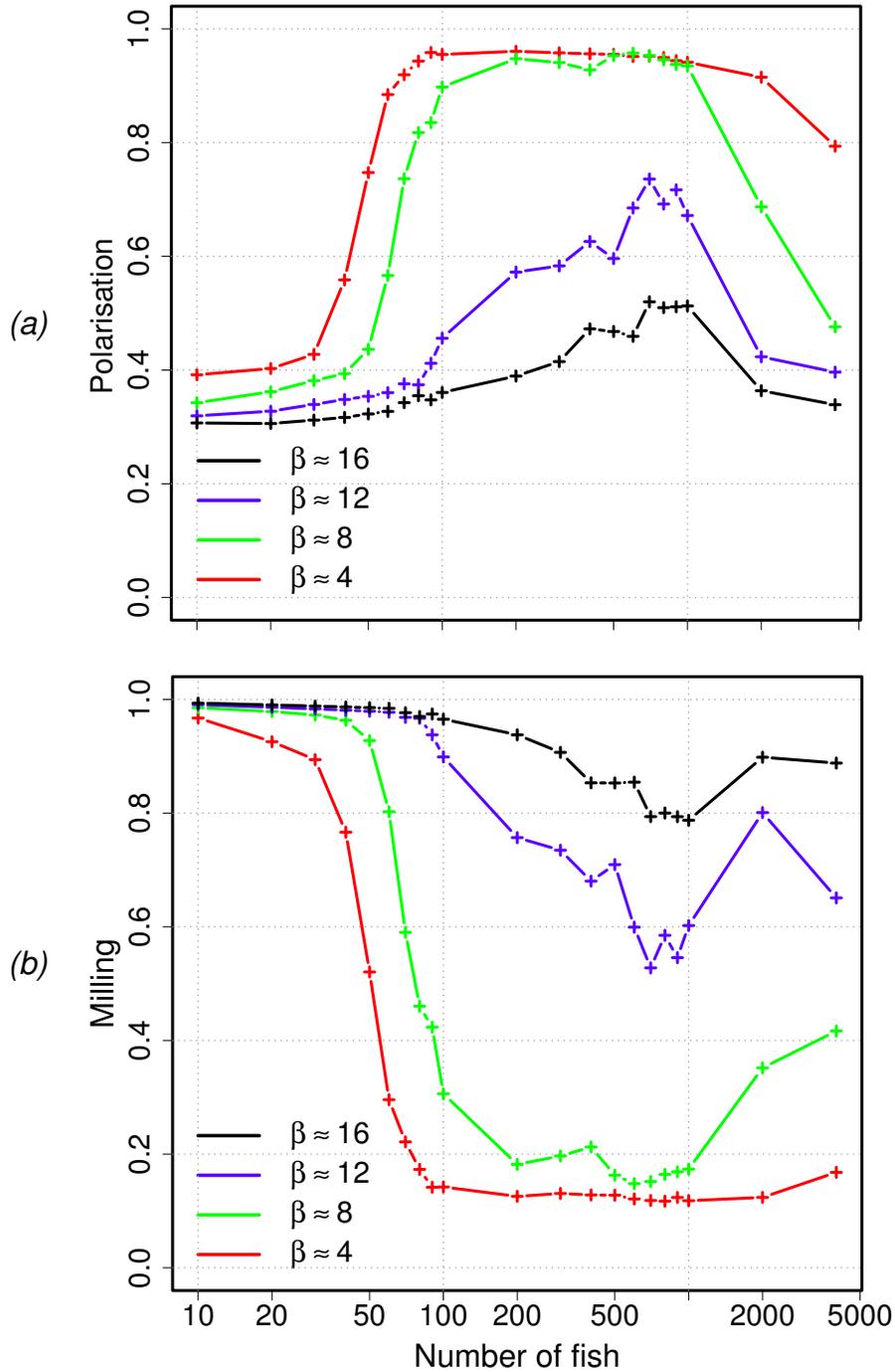}\
 \caption{Simulations done for a fixed $\gamma \approx 4.5$ with an
  angular inertia term of $\alpha \approx 0.014$ (equivalent to a speed
  $v=0.8 {\rm m/s}$) for different quantities of fish and $\beta$ values, where
  $(a)$ and $(b)$ refer to the milling and polarisation parameter
  respectively.}
\label{Size}
\end{center}
\end{figure}

We performed a series of simulations at fixed $\gamma \approx 4.5$
with an angular inertia term $\alpha \approx 0.014$ (equivalent to a
speed $v=0.8 {\rm m/s}$). Figure \ref{Size} shows the milling and
schooling order parameters for simulations of groups from 10 to 4000
fish.  These results indicate that there exists an inferior limit of
approximately 60 fish in order to achieve significant milling and that
this behaviour progressively disappears at large sizes.  Although
these results must be taken with care given the unbounded character of
the variation of positional interaction with distance to neighbours,
they show that for up to 100 fish, increasing group size induces a
transition from schooling to milling for $\beta \le 8$. For larger
values of the alignment parameter, despite having higher overall
values for the milling parameter, the transition does not occur. This
indicates that in the model studied here, like in many of those
proposed before \cite{Lukeman2009}, milling dynamics does not emerge
in the arbitrarily large infinite-size limit.

Although such a statement has obvious interest for physicists, we
believe it bears some importance regarding animal group
behaviour. Also, it is worth noting that the variation on the milling
parameter seen for $\beta \ge 12$ in figure \ref{Size}$(a)$ happens as
the duration of the milling and schooling behaviour increases with
$N$. Such an increase is so intense that for simulations with more
than 300 fish, usually only one behaviour can be seen for every
initial condition, giving a very large standard deviation and a
highly fluctuating average. Furthermore, some simulations for $N>1000$
displayed more than one vortex at the same time, patterns which the
milling parameter cannot account for. One can see in figure
\ref{Size}$(b)$ that the schooling behaviour is affected by the size
as well. This again is due to the effect of large schools. This
stretch on school extensions enables information propagation delays,
which in turn, cause reductions on the global polarisation
parameter. It is easy to see how a three dimensional configuration
could achieve lower extensions with large fish quantities, which would
in turn minimise these effects.

\section{Conclusions}
\label{SecConc}

Understanding how complex motion patterns in fish schools arise from
local interactions among individuals is a key question in the study of
collective behaviour \cite{Lopez2012, Delcourt2012}. In a previous
work, Gautrais \etal have determined the stimulus/response function
that governs an individual’s moving decisions in Barred flagtail
({\it Kuhlia mugil}) \cite{Gautrais2012}. It has been shown that two
kinds of interactions controlling the attraction and the alignment of
fish are involved and that they are weighted continuously depending on the position and orientation of the neighboring fish.
It has also been found that the magnitude of these
interactions changes as a function of the swimming speed of fish and
the group size. The consequence being that groups of fish adopt
different shapes and motions: group polarisation increases with
swimming speed while it decreases as group size increases.
 
Here we have shown that the relative weights of the attraction and alignment
interactions play a key role in the emergent collective states at the
school level. Depending on the magnitude of the attraction and the
alignment of fish to their neighbours, different collective states can
be reached by the school. The exploration of the parameter space of
the Gautrais {\it et al} model reveals the existence of two
dynamically stable collective states: a swarming state in which
individuals aggregate without cohesion, with a low level of
polarisation, and a schooling state in which individuals are aligned
with each others and with a high level of polarisation. The transition
between the two states is induced by an increase of the swimming
speed. Furthermore, the addition in the model of a frontal preference
to account for the angular weighting of interactions leads to two other
collective states: a milling state in which individuals constantly
rotate around an empty core thus creating a torus and a winding state,
in which the group self-organises into a linear crawling
structure. This last group structure is reminiscent of some moving
patterns observed in the Atlantic herring ({\it Clupea harengus})
\cite{Gerlotto2003}.
 
Of particular importance is the transition region between milling and
schooling. In this region, the school exhibits multistability and it
regularly shifts from schooling to milling for the same combination of
individuals' parameters. This particular property was recently
reported in experiments performed in groups of golden shiners
\cite{Tunstrom2013}. Our results show that the transition region can
be described by a simple functional form describing the respective
weights of the alignment and attraction parameters and is independent
of the fish swimming speed in the experimental range. The modulation
of the strength of the alignment and attraction may depend on the
behavioural and physiological state of fish \cite{Wendelaar1997}. In
particular, various environmental factors such as a perceived threat
may change the way fish respond to their neighbours and hence lead to
dramatic changes in collective motion patterns at the school level
\cite{Beecham1999, Viscido2002}.
 
Finally, we show that collective motion may dramatically change as
group size increases. The absence of a milling state in the largest
groups is a natural consequence of the spatial constraints exerted
upon individuals' movements as the number of fish exceeds some
critical value. These constraints could be much less stringent in 3D,
as testified by the observation of cylindrical vertical milling structures involving
several thousands of fish in bigeye trevally ({\it Caranx
  sexfasciatus}). With an additional dimension, the same number of
fish could result into a mill of much smaller diameter, not only avoiding the
problems of distance dependence, but also minimizing information
propagation delays and the emergence of competing behaviours.

A more in depth physics study of the transitions presented here is
necessary. Unfortunately the present model is restricted to schools of
biologically-relevant sizes ($N \approx 100$ fish) preventing such
analysis for the current social interactions. In this manner,
additional changes to the model are required, among them are: i)
eliminating the linear distance dependence on the interactions, by
either implementing a saturation or a decay to this interaction; ii)
changing the boundary conditions for periodic ones to avoid
evaporating schools once the distance dependence is removed; iii) a
three dimensional study of the model to check the impact of an
additional level of freedom on the school behaviour.

\section{Acknowledgements}

We are grateful to Jacques Gautrais and Sepideh Bazazi for comments on
the manuscript. Daniel S. Calovi was funded by the Conselho Nacional
de Desenvolvimento Cient{\'i}fico e Tecnol{\'o}gico - Brazil. Ugo
Lopez was supported by a doctoral fellowship from the scientific
council of the Universit{\'e} Paul Sabatier. This study was supported
by grants from the Centre National de la Recherche Scientifique and
Universit{\'e} Paul Sabatier (project Dynabanc).

\section{References}

\bibliography{references.abv} \bibliographystyle{unsrt}

\section{Figures}

\begin{figure}[!htb]
\begin{center}
\includegraphics[width=0.9\linewidth]{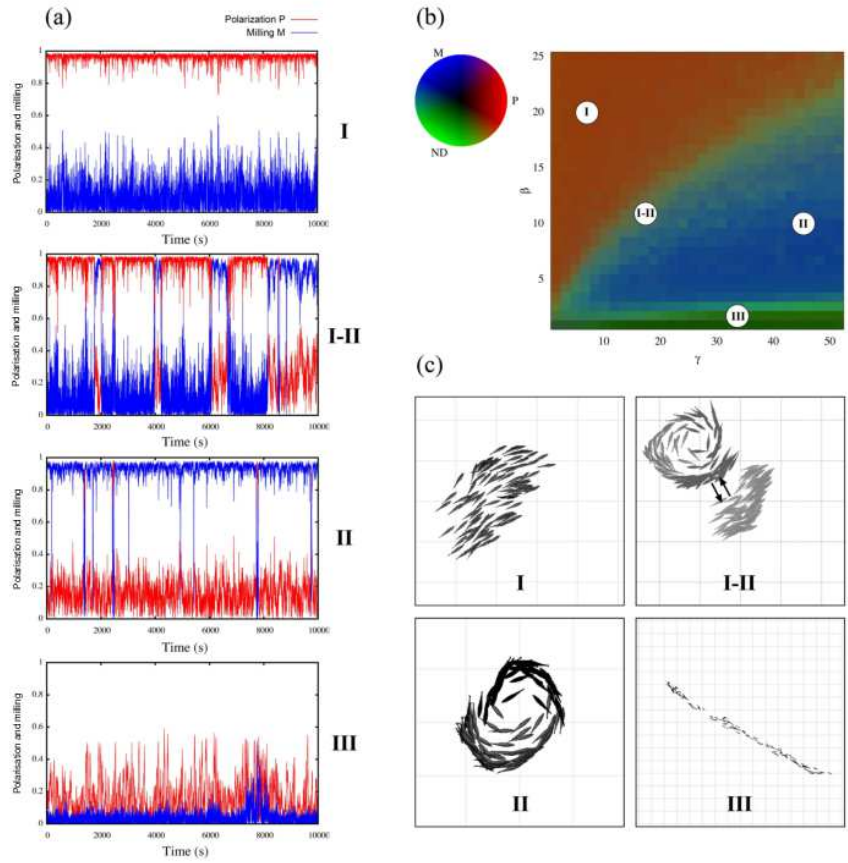}

\caption{ $(a)$ The respective time series of the polarisation (red
  line) and milling parameters (blue line) for the four distinct
  regions.  $(b)$ is a representation of \fref{v04F0F2} right panels
  where distinct behavioural regions can be identified.  $(c)$ typical
  configurations of these states.}
\label{Phases}
\end{center}
\end{figure}

\begin{figure}[!htb]
\begin{center}
\includegraphics[width=0.9\linewidth]{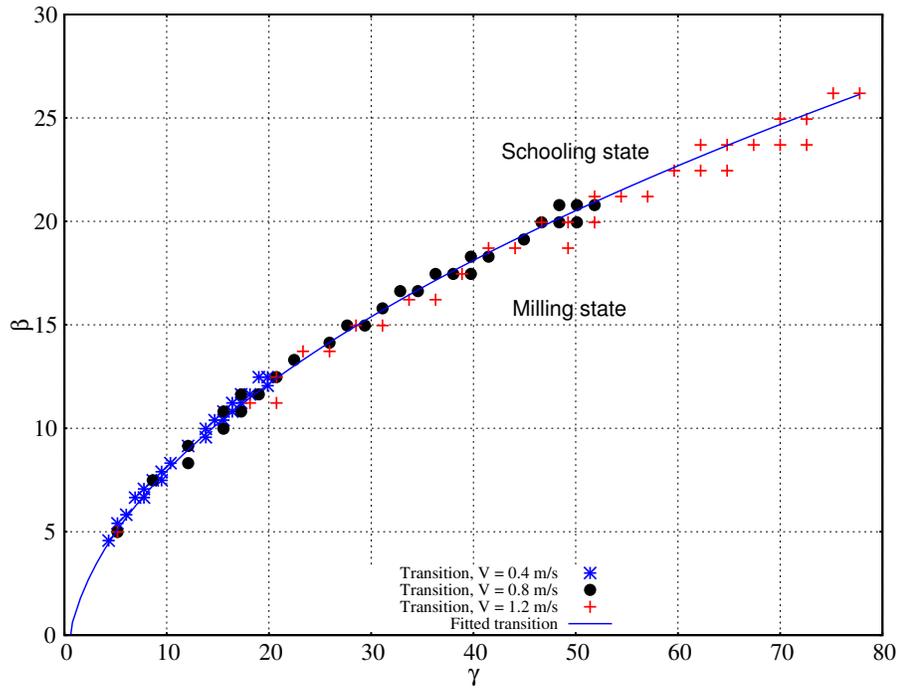}
 \caption{Transition between region schooling and milling states for
   different speeds.  A functional form ($A\sqrt{\gamma}+B$) fitted for
   $v$ = 0.8 m/s proved to also properly describe the transitions
   for speeds of 0.4 and 1.2 m/s, where $A=3.22\pm 0.05$ and $B=-2.23
   \pm 0.26$.
in order to better}
\label{Trans}
\end{center}

\end{figure}

\clearpage

\appendix
\section{Supplementary information} 

\begin{sup}
\begin{center}
\caption{Region\_I.avi - Simulation with 100 fish made on the
  schooling parameter space region as seen in \fref{Phases}$(b)$. Note
  the data was post-treated so that the simulations seem to have been
  made with periodic boundary conditions. This representation was
  chosen in order to have an easier visualisation of the group
  dynamics, but all simulations were performed in an infinite spatial
  domain.}
\label{RegionI}
\end{center}
\end{sup}

\begin{sup}
\begin{center}
\caption{Region\_II.avi - Simulation with 100 fish made on the milling
  parameter space region as seen in \fref{Phases}$(b)$. Note the data
  was post-treated so that the simulations seem to have been made with
  periodic boundary conditions. This representation was chosen in
  order to have an easier visualisation of the group dynamics, but all
  simulations were performed in an infinite spatial domain.}
\label{RegionII}
\end{center}
\end{sup}

\begin{sup}
\begin{center}
\caption{Region\_I-II.avi - Simulation with 100 fish made on the
  transition between schooling and milling parameter space region as
  seen in \fref{Phases}$(b)$. Note the data was post-treated so that
  the simulations seem to have been made with periodic boundary
  conditions. This representation was chosen in order to have an
  easier visualisation of the group dynamics, but all simulations were
  performed in an infinite spatial domain.}
\label{RegionI-II}
\end{center}
\end{sup}

\begin{sup}
\begin{center}
\caption{Region\_III.avi - Simulation with 100 fish made on the line
  parameter space region as seen in \fref{Phases}$(b)$. Note the data
  was post-treated so that the simulations seem to have been made with
  periodic boundary conditions. This representation was chosen in
  order to have an easier visualisation of the group dynamics, but all
  simulations were performed in an infinite spatial domain.}
\label{RegionIII}
\end{center}
\end{sup}

\begin{sup}[ht]
\begin{center}
\caption{Gamma.avi - Two dimensional histogram evolution for the schooling and
  milling parameters as we change $\beta$ while maintaining $
  \gamma=17.28$, meaning a vertical cross section on the transition
  represented on \fref{Trans}.}
\label{gamma}
\end{center}
\end{sup}

\begin{sup}[ht]
\begin{center}
\caption{Beta.avi - Two dimensional histogram evolution for the
  schooling and milling parameters as we change $\gamma$ while
  maintaining $ \beta=13.30$, meaning a horizontal cross-section on
  the transition represented on \fref{Trans}.}
\label{beta}
\end{center}
\end{sup}

\end{document}